\documentclass{PoS}
\pdfpageattr {/Group << /S /Transparency /I true /CS /DeviceRGB>>}
\usepackage{wrapfig}

\title{The spin structure of the proton at low $x$ and low $Q^2$ in two-dimensional bins from COMPASS}

\ShortTitle{The spin structure of the proton at low $x$ and low $Q^2$ in two-dimensional bins from COMPASS}

\author{\speaker{A.S.~Nunes}
         \thanks{Financed by FCT, grant CERN/FIS-NUC/0017/2015.}\\
        LIP, Lisbon\\
        E-mail: \email{ana.sofia.nunes@cern.ch}}
\author{on behalf of the COMPASS Collaboration}

\abstract{The longitudinal double spin asymmetries $A_1^p$ and the spin dependent structure function of the proton $g_1^p$ were extracted from COMPASS data in the region of low Bjorken scaling variable $x$ and low photon virtuality $Q^2$. The data were taken in 2007 and 2011 from scattering of polarised muons off polarised protons, resulting in a sample that is 150 times larger than the one from the previous experiment SMC that pioneered studies in this kinematic region.

For the first time, $A_1^p$ and $g_1^p$ were evaluated in this region in two-dimensional bins of kinematic variables: $(x,Q^2)$, $(\nu ,Q^2)$, $(x,\nu)$ and $(Q^2,x)$. The following kinematic region was investigated: $4\times 10^{-5}<x<4\times 10^{-2}$, $0.001$~(GeV/$c$)$^2<Q^2<1$~(GeV/$c$)$^2$ and $14$~GeV$<\nu <194$~GeV. The obtained results were confronted with theoretical models.}

\FullConference{XXIV International Workshop on Deep-Inelastic Scattering and Related Subjects\\
                 11-15 April, 2016\\
                 DESY Hamburg, Germany}

\begin{document}

\section{Introduction}
The inelastic scattering of polarised leptons off polarised nucleons has played an important role in the study of the structure of nucleons in the last few decades. Particularly interesting is the region of low $x$, corresponding to high parton densities. However, in fixed target experiments, $x$ and $Q^2$ are highly correlated, and probing the low $x$ region implies also entering the non-perturbative region of $Q^2<1$~(GeV/$c$)$^2$, which is poorly known experimentally and where perturbative QCD cannot be applied. Nevertheless, this region allows studies of the transition from the non-perturbative regime of photoproduction to the perturbative region of deep inelastic scattering where several predictions were formulated~\cite{bkz2002,egt2008,egt2010}. Previously, the Spin Muon Collaboration (SMC) has presented its results on $A_1^p$ and $g_1^p$ in this region~\cite{SMClowx}. However, large errors have not permitted a detailed comparison with predictions. COMPASS has a large sample in this kinematic region, of the order of $7\times 10^8$~events, and it is now possible to extract $A_1^p$ and $g_1^p$ in two-dimensional bins of the following variables: $(x,Q^2)$, $(\nu,Q^2)$, $(x,\nu)$ and $(Q^2,x)$, where $\nu$ is the energy difference between the incoming and the scattering lepton.

\section{Method}
The COMPASS collaboration runs a fixed target experiment at the CERN SPS~\cite{COMPASSnima}. It uses a multipurpose apparatus, and only the setup relevant for the measurement described here will be detailed. The experiment uses a tertiary muon beam that is naturally polarised. In 2007 and 2011, respectively, a positive muon beam of 160 GeV or 200 GeV with a longitudinal polarisation of around 80\% was used. The beam impinges on a 1.2~m long solid state target of ammonia that contains polarised protons with a polarisation of about 85\%. The dilution factor, which accounts for the fraction of the material that is polarisable in the target is about 16\%. The target polarisation is built up and maintained using a superconducting solenoid providing a magnetic field of up to 2.5~T and a dilution refrigerator that allows temperatures of the target material as low as 60~mK. A large acceptance, two-staged spectrometer contains two dipole magnets and tracking, calorimetry and particle identification detectors in both stages of the setup.

During the 2007 and the 2011 data taking periods, the target was divided in tree cells, of 30, 60 and 30 cm, with consecutive cells longitudinally polarised in opposite directions. This allowed to simultaneously record data for the two target spin configurations, with similar acceptances for both. In order to further reduce possible systematic effects, the polarisation directions were rotated, typically every twenty-four hours. In addition, the relative directions of target cell polarisations with respect to the target solenoid field direction were swapped at least once per year, to further minimise possible systematic uncertainties.

\begin{figure}
\centering
\begin{tabular}{cc}

\includegraphics[width=0.42\textwidth]{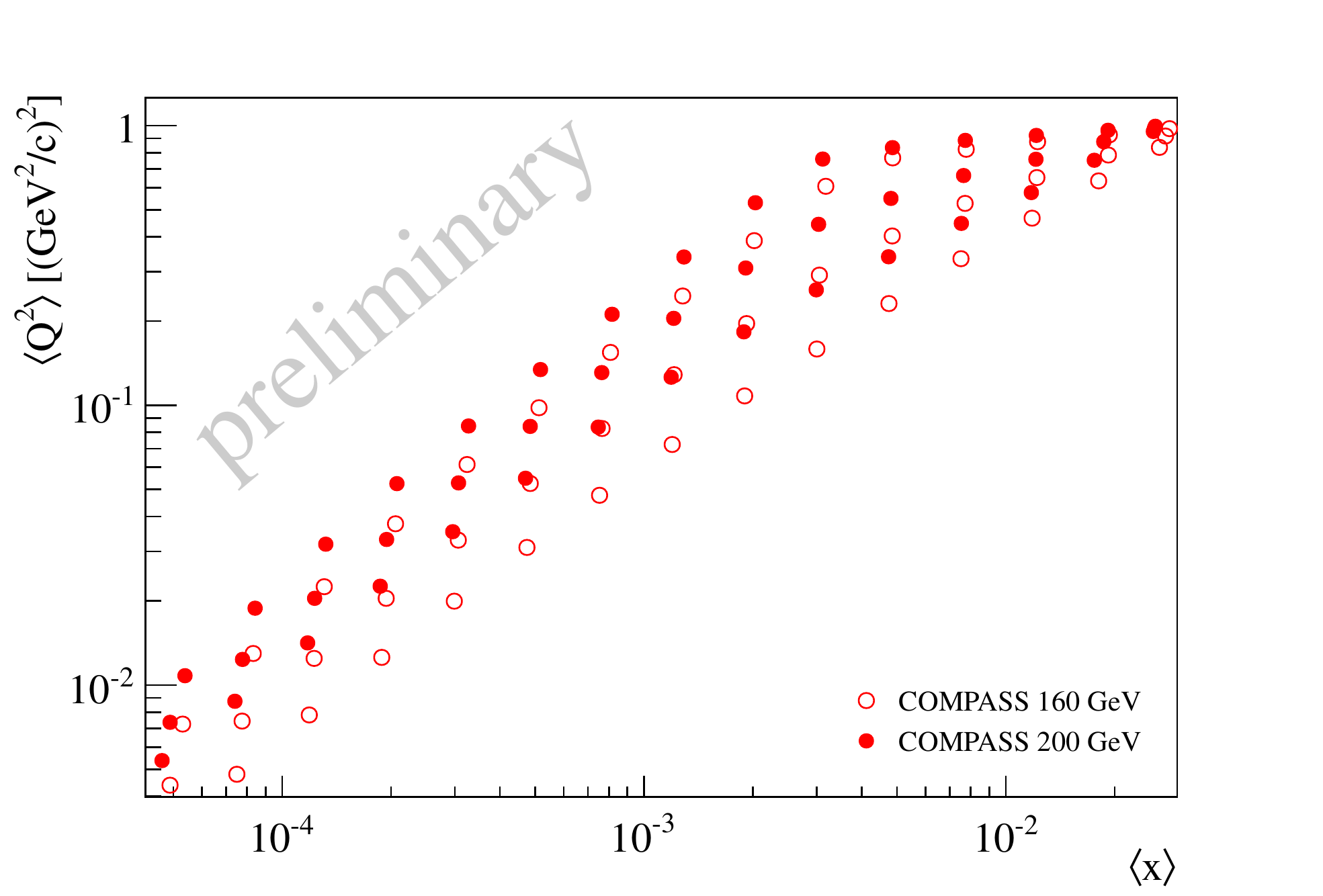}&
\includegraphics[width=0.42\textwidth]{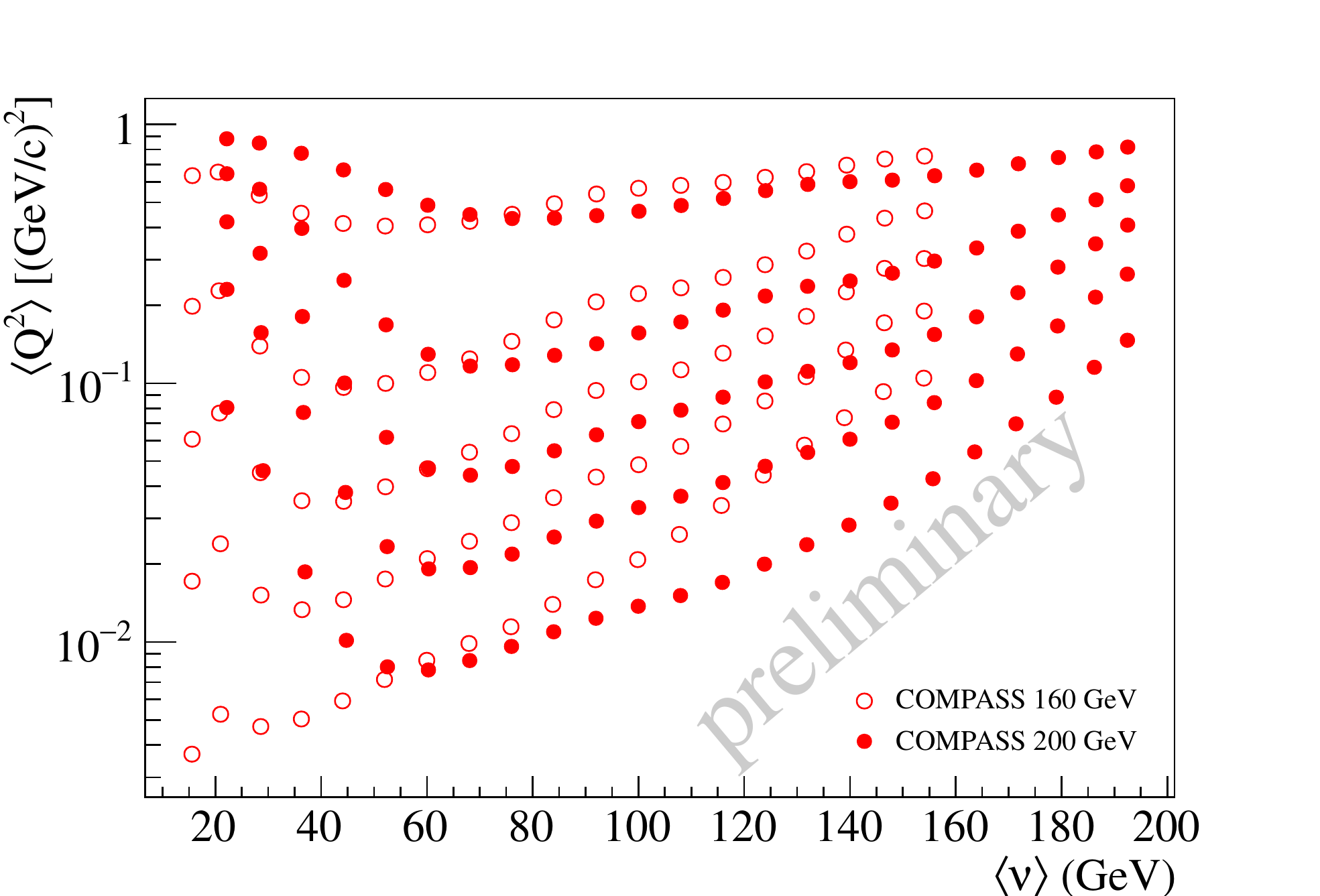}\\ 
(a)&(b)\\                 
\includegraphics[width=0.42\textwidth]{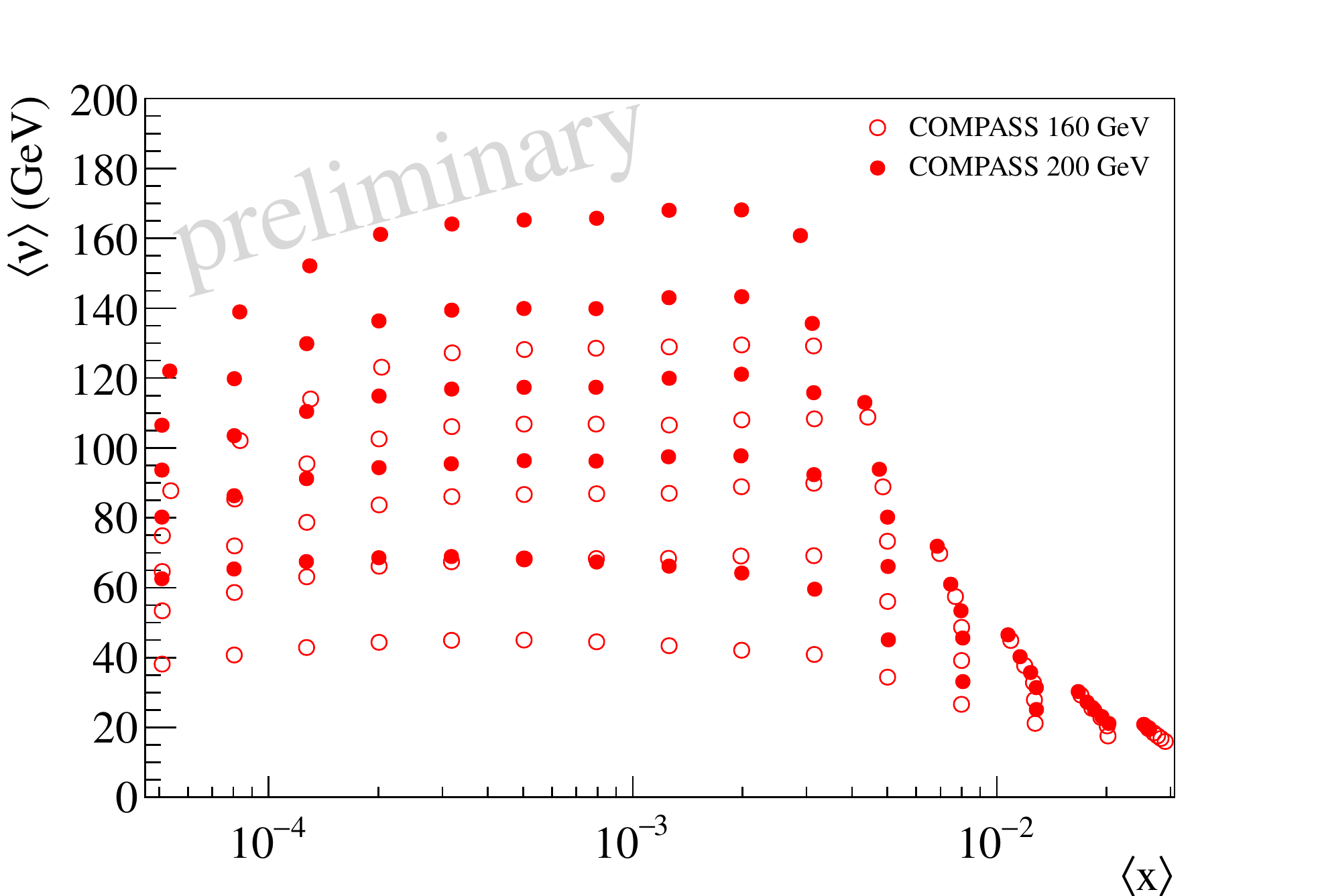}&
\includegraphics[width=0.42\textwidth]{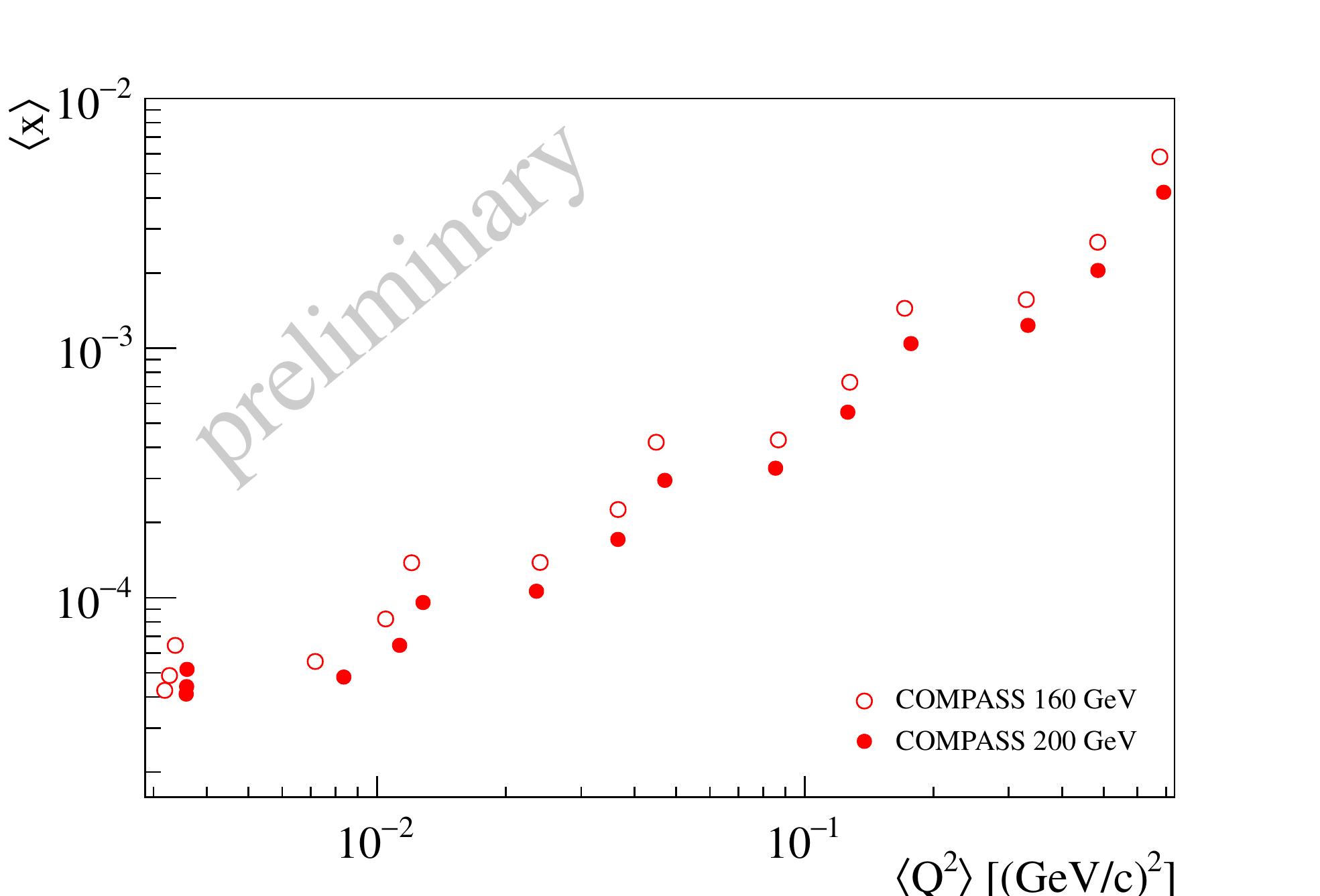}\\
(c)&(d)\\
\end{tabular}
\caption{Average values of the pairs of independent variables for each of the four sets of two-dimensional bins used: (a) $(x,Q^2)$, (b) $(\nu,Q^2)$, (c) $(x,\nu)$ and (d) $(Q^2,x)$.}
\label{fig:bins}
\end{figure}

The main selection criteria of the events include $Q^2<1$~(GeV/$c$)$^2$, $x\geq 4\times 10^{-5}$ and $0.1<y<0.9$, where $y$ is the fraction of energy loss by the muon in the nucleon's rest frame. Due to the low scattering angle of the scattered muon, the position of the interaction inside the target cannot be determined precisely. Therefore, at least one additional track besides the scattered muon must originate from the interaction point. This ``hadron method'' was first introduced by the SMC experiment which also showed that inclusive asymmetries at low $x$ are not biased by this method~\cite{F2pSmcFit}. It is also required that the selected events do not come from the elastic scattering of beam muons off target material electrons.
The final asymmetry thus obtained consists of $676\times 10^6$ events, of which $447\times 10^6$ were collected in 2007 with a 160 GeV beam and $229\times 10^6$ events were collected in 2011 with a 200 GeV beam. This is about a factor 150 times larger than the data sample of the SMC in a similar phase-space region. The data were divided into bins of two kinematic variables, whose average values are shown in Fig.~\ref{fig:bins} for the four sets of two-dimensional bins used.

The number of events collected of the two relative spin configurations (of parallel and antiparallel beam and target polarisations) is related to the double longitudinal spin asymmetry of the proton $A_1^p$ by
\begin{equation}
N^{
\stackrel{\leftarrow}{\Rightarrow}
,
\stackrel{\leftarrow}{\Leftarrow}
}\simeq a\,\phi \, n\,\bar{\sigma }(1\pm P_b\, P_t\, f\, D\, A_1^p).
\label{eq:nevents}
\end{equation}
Here $a$ is the acceptance, $\phi$ is the beam flux, $n$ is the number of nucleons in the target, $\bar{\sigma}$ is the muon-proton spin-independent cross-section, $P_b$ and $P_t$ are, respectively, the beam and target polarisations, $f$ is the dilution factor and $D$ is the depolarisation factor, that accounts for the fraction of the muon polarisation that is transfered to the virtual photon.

Data are taken before and after field rotations that invert the target cell polarisations. Thus the ratio 
$(N^{\stackrel{\leftarrow}{\Rightarrow},1}\cdot
N^{\stackrel{\leftarrow}{\Rightarrow},2})/(N^{\stackrel{\leftarrow}{\Leftarrow},1}\cdot
N^{\stackrel{\leftarrow}{\Leftarrow},2})$
can be directly computed; the superscripts 1 and 2 refer to before and after the field rotation. After substituting the event number expressions from Eq.~\ref{eq:nevents} and making reasonable assumptions on the stability of the experimental setup, we obtain a second order equation for $A_1^p$ that can be solved to obtain this quantity. To optimise the statistical errors, each event is given a weight $\omega = f\, D\, P_b$. This is done for sets of data of the two relative spin configurations obtained close in time, typically every forty-eighty hours. These asymmetries per configuration are combined into weighted averages of final asymmetries for each kinematic bin. To account for spin-independent radiative events, the program TERAD~\cite{terad} is used. The corrections, which are functions of $x$ and $y$ (the fraction of the lepton's energy lost in the nucleon rest frame), are included in the (effective) dilution factor. Spin-dependent corrections are obtained using the program POLRAD~\cite{polrad} and are applied to the asymmetries. The asymmetries were further corrected for the presence of polarisable $^{14}$N in the target material. Thorough checks were done to identify possible sources of false asymmetries, leading to the conclusion that the systematic uncertainties on asymmetries are similar in size to the statistical ones.

The spin dependent structure function $g_1^p$ is obtained from $A_1^p$ using
\begin{equation}
g_1^p=\frac{F_2^p}{2\, x\, (1+R)}A_1^p.
\label{eq:g1}
\end{equation}
Here $F_2^p$ is the spin independent structure function, which comes either from the SMC fit to data~\cite{F2pSmcFit}, or, in the case of low $x$ and low $Q^2$, from a model~\cite{F2pModel}. $R$ is the ratio of the absortion cross-sections of the longitudinally and transversely polarised virtual photon, which is based on the SLAC parameterisation~\cite{SLACparam} extended to low $Q^2$ values~\cite{COMPASSdeuteronQ2lt1}.

\section{Results and discussion}

\begin{wrapfigure}{r}{0.5\textwidth}
\centering\includegraphics[width=0.56\textwidth]{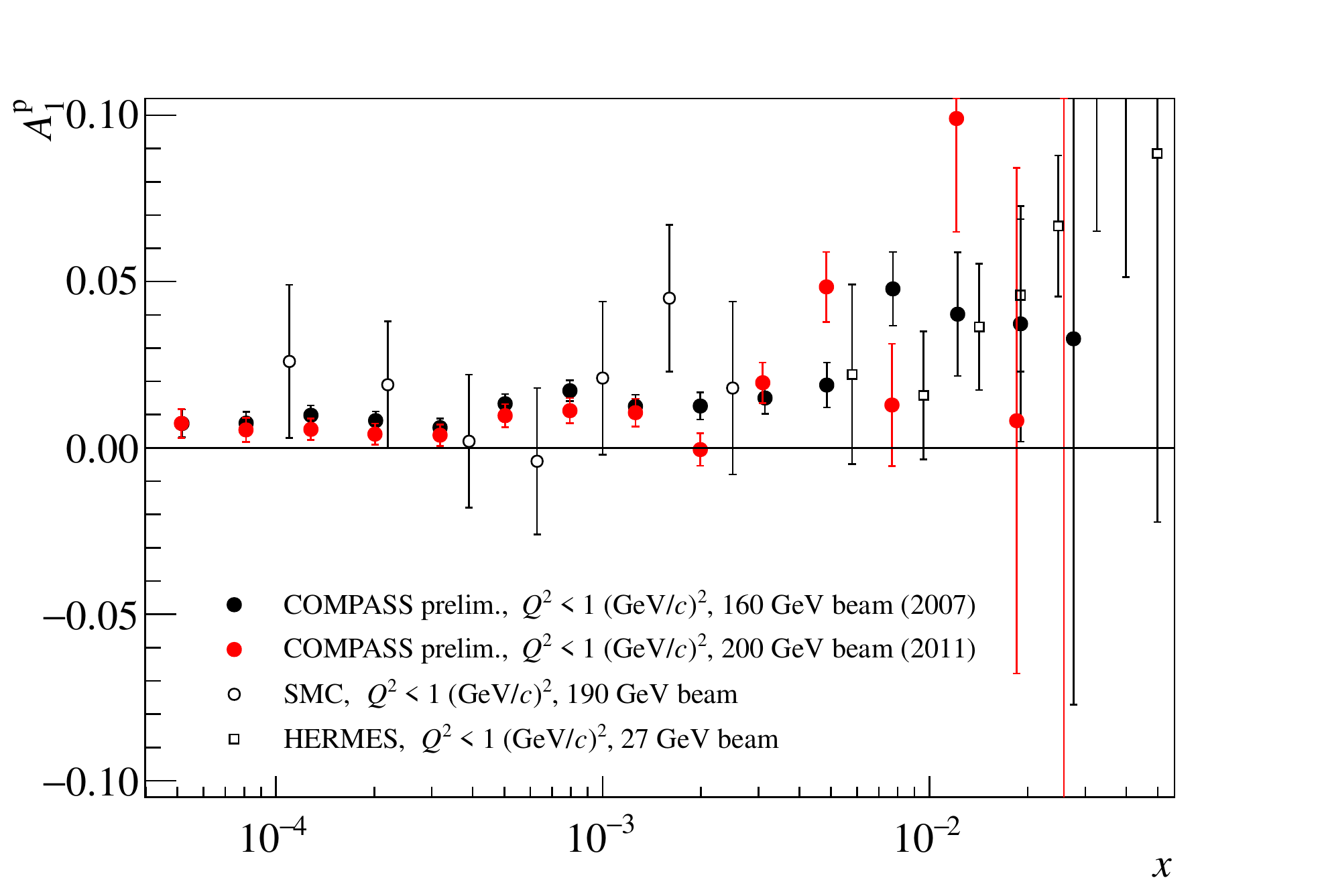}
\caption{$A_1^p(x)$ at low $x$ and low $Q^2$ from COMPASS and from previous experiments~\cite{SMClowx,SMCgeneric,HERMESdata}. 
A clear positive asymmetry is seen at very low $x$ of the COMPASS data
which also demonstrate a significantly improved precision of the
measurement.
}
\label{fig:A1}
\end{wrapfigure}

The longitudinal double spin asymmetries at low $x$ for the proton observed by COMPASS are positive, see Fig.~\ref{fig:A1}, at contrast with the results of the SMC, albeit within large errors of the latter. COMPASS asymmetries for the deuteron in this region are zero~\cite{COMPASSdeuteronQ2lt1}. No particular trends are visible for the asymmetries nor for the structure function $g_1^p$ as functions of $x$ and $\nu$. 

\begin{figure}
\begin{tabular}{cc}
\raisebox{14pt}{\includegraphics[width=0.545\textwidth]{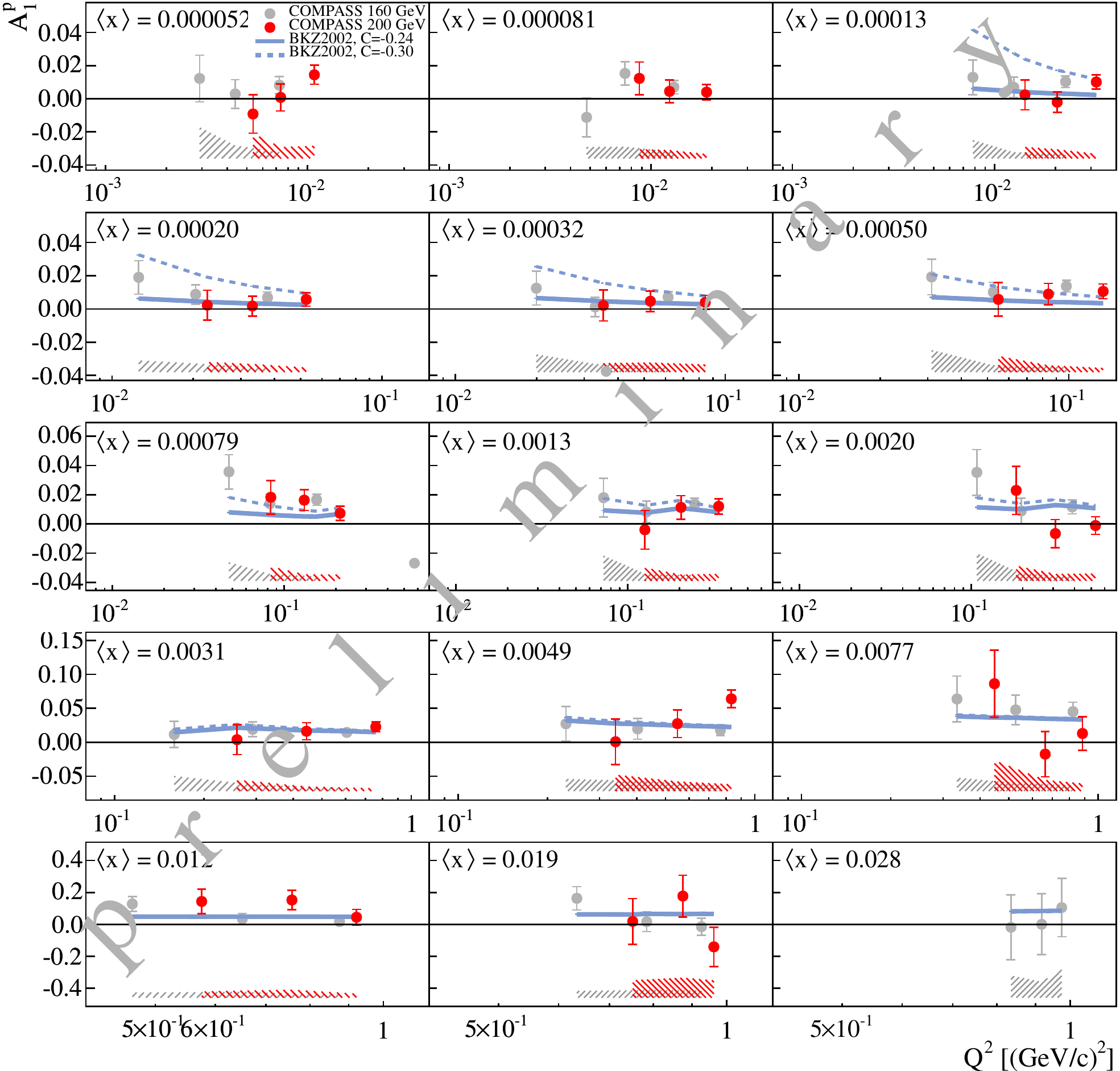}}&
\includegraphics[width=0.455\textwidth]{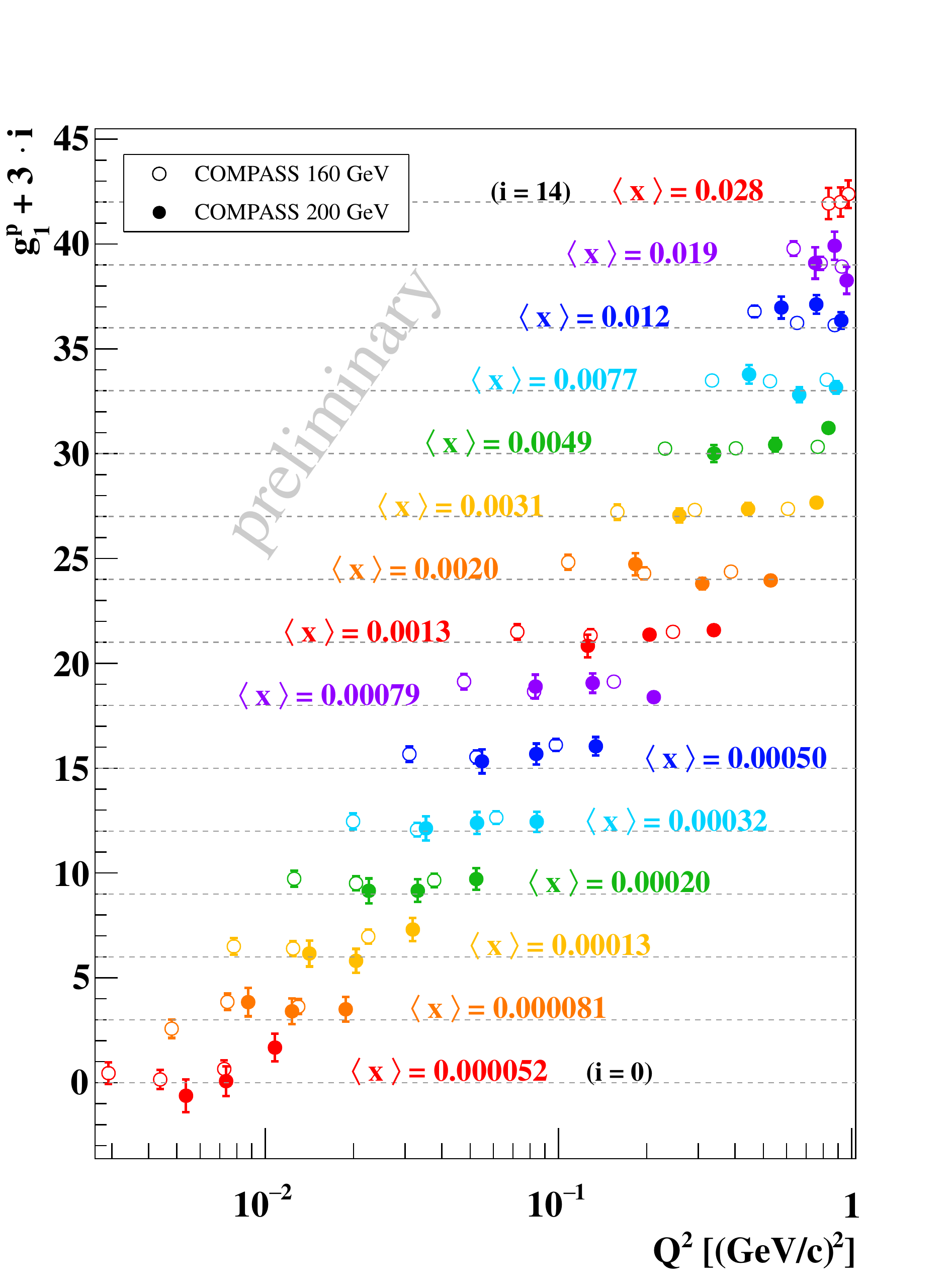}\\
(a)&(b)\\
\end{tabular}
\caption{
The asymmetry $A_1^p(Q^2,x)$ (a) and structure function $g_1^p(Q^2,x)$ (b) extracted from the 2007 and 2011 data. Lines in Fig.~(a) mark predictions
of the model~\cite{bkz2002}. 
}
\label{fig:A1g1}
\end{figure}

The $A_1^p$ and $g_1^p$ extracted in the first two-dimentional set of bins $(x,Q^2)$ is shown in Fig.~\ref{fig:A1g1}. Superimposed are the predictions of the model from Ref.~\cite{bkz2002}, based on GVMD ideas. We can see slightly positive asymmetries, as in the one-dimensional analyses.

There are no significant differences between the datasets obtained with two diferent beam energies, and there is no significant trend in the data as function of $Q^2$ or $x$.
There is a reasonable com\-pa\-ti\-bi\-li\-ty with the predictions of the model~\cite{bkz2002}. The same conclusions can be drawn for the other three sets of two-dimensional bins studied, $(\nu,Q^2)$, $(x,\nu)$ and $(Q^2,x)$. 
Furthermore, no strong dependence of $g_1^p$ with $x$ or with $Q^2$ is visible in Fig.~\ref{fig:A1g1}(b).

\end{document}